\documentclass[12pt]{iopart}
\expandafter\let\csname equation*\endcsname\relax
\expandafter\let\csname endequation*\endcsname\relax
\usepackage{amsmath}
\usepackage{graphicx}
\usepackage{bm}
\usepackage{amssymb}
\usepackage{dcolumn}
\usepackage{subfigure}
\usepackage{threeparttable}
\usepackage{multirow}
\usepackage{mathrsfs}
\DeclareMathAlphabet{\mathscrbf}{OMS}{mdugm}{b}{n}
\usepackage[usenames,dvipsnames]{color}
\usepackage{mathtools}

\begin{document}
\newcommand{\vn}[1]{{\boldsymbol{#1}}}
\newcommand{\vht}[1]{{\boldsymbol{#1}}}
\newcommand{\matn}[1]{{\bf{#1}}}
\newcommand{\matnht}[1]{{\boldsymbol{#1}}}
\newcommand{\bege}{\begin{equation}}
\newcommand{\ee}{\end{equation}}
\newcommand{\bal}{\begin{aligned}}
\newcommand{\defbar}{\overline}
\newcommand{\SM}{\scriptstyle}
\newcommand{\eal}{\end{aligned}}
\newcommand{\udot}{\overset{.}{u}}
\newcommand{\exponential}[1]{{\exp(#1)}}
\newcommand{\phandot}[1]{\overset{\phantom{.}}{#1}}
\newcommand{\phandag}{\phantom{\dagger}}
\newcommand{\Trace}{\text{Tr}}
\newcommand{\Bxc}{\Omega}
\newcommand{\Torque}{\tau}
\setcounter{secnumdepth}{2}
\title[The inverse thermal spin-orbit torque and its relation to DMI]{
The inverse thermal spin-orbit torque and the
relation of the Dzyaloshinskii-Moriya interaction
to ground-state energy currents}
\author{Frank Freimuth, Stefan Bl\"ugel and Yuriy Mokrousov}
\address{Peter Gr\"unberg Institut and Institute for
Advanced Simulation,
Forschungszentrum J\"ulich and JARA, 52425 J\"ulich, Germany}
\ead{f.freimuth@fz-juelich.de}
\begin{abstract}
Using the Kubo linear-response formalism
we derive expressions to calculate the
heat current generated by magnetization dynamics in magnets
with broken inversion symmetry and spin-orbit interaction (SOI).
The effect of producing heat currents by magnetization dynamics constitutes
the Onsager reciprocal of the thermal spin-orbit torque (TSOT), i.e.,  
the generation of torques on the magnetization due to temperature gradients.
We find that the energy current driven by magnetization
dynamics contains a contribution from the 
Dzyaloshinskii-Moriya interaction (DMI), which needs to be
subtracted from the Kubo linear response of the energy current
in order to extract the heat current.
We show that the expressions of the DMI coefficient can be 
derived elegantly from the DMI energy current. 
Guided by formal analogies between the Berry phase theory of DMI on
the one hand and
the modern theory of orbital magnetization on the other hand 
we are led to an interpretation of the latter in terms of energy
currents as well. Based on \textit{ab-initio} calculations we
investigate the heat current driven by magnetization dynamics in
Mn/W(001) magnetic bilayers. We predict that fast domain walls drive
strong ITSOT heat currents.
\end{abstract}

\pacs{72.25.Ba, 72.25.Mk, 71.70.Ej, 75.70.Tj}


\maketitle
\section{Introduction}
\newcommand{\myfigurewidth}{0.82\linewidth}
The interaction of heat current with electron spins is at the
heart of spin caloritronics~\cite{spincat_review_Bauer}.
It leads to thermal spin-transfer torques (STTs) on the magnetization in
spin valves, magnetic tunnel junctions, and domain 
walls when a temperature gradient is applied~\cite{thermally_driven_dwm_fe_w110,thermal_stt_yuan,landau_lifshitz_theory_thermomagnonic_torque,thermal_stt_hatami,thermal_stt_femgofe_jia,evidence_thermal_stt,parameter_space_thermal_stt}.
While the thermal STT does not require spin-orbit interaction (SOI), it
only exists in noncollinear magnets. In spin valves and 
magnetic tunnel junctions this noncollinearity arises when the 
magnetizations of the free and fixed layers are not parallel,
while in domain walls it arises from the continuous rotation of
magnetization across the wall.

In the presence of SOI electric currents and heat currents can generate torques also in
collinear magnets:
In ferromagnets with broken inversion symmetry
the so-called
spin-orbit torque (SOT) acts on the
magnetization when an electric current 
is applied (Figure~1a)~\cite{manchon_zhang_2008,torque_macdonald,
chernyshov_2009,current_induced_switching_using_spin_torque_from_spin_hall_buhrman,
symmetry_spin_orbit_torques,
semi_classical_modelling_stiles,antidamping_kurebayashi,ibcsoit,rashba_review}
.
The inverse spin-orbit torque (ISOT)
consists in the production of an electric current due to
magnetization dynamics (Figure~1b)~\cite{invsot,
spin_motive_force_hals_brataas,charge_pumping_Ciccarelli}.
The application of a temperature gradient
results in the thermal spin-orbit 
torque (TSOT) (Figure~1c)~\cite{fept_guillaume}.
TSOT and SOT are related by a Mott-like expression~\cite{mothedmisot}.

\begin{figure}
\flushright
\includegraphics[width=\myfigurewidth]{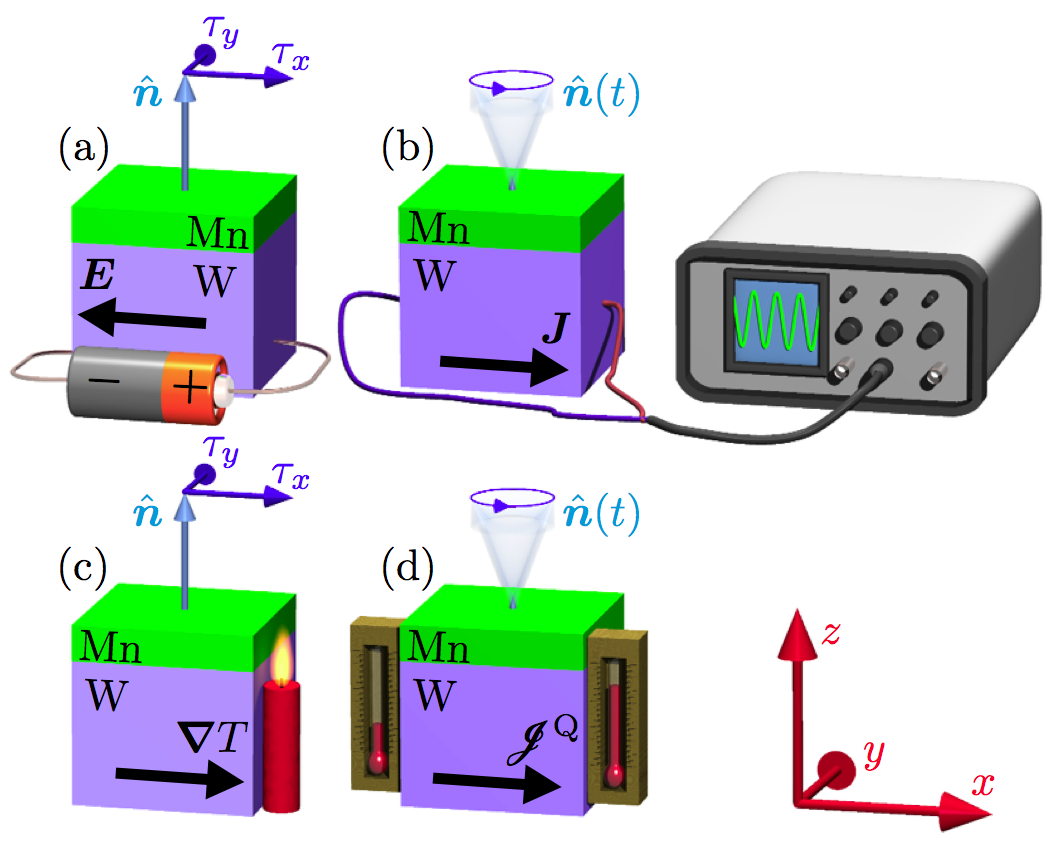}
\caption{\label{figuresotfamily}Family of 
SOT-related effects in a Mn/W magnetic bilayer with
broken structural inversion symmetry. 
(a) SOT: An applied electric field $\vn{E}$ generates a torque $\vn{\Torque}$
on the magnetization. $\hat{\vn{n}}$ is the magnetization direction.
(b) ISOT: Magnetization dynamics $\partial \hat{\vn{n}}/\partial t$ 
drives an electric current $\vn{J}$.
(c)~TSOT: The application of a temperature gradient $\vn{\nabla}T$
generates a torque $\vn{\Torque}$.
(d) ITSOT: Magnetization dynamics drives a 
heat current $\mathscrbf{J}^{\rm Q}$.
}
\end{figure}

In this work we discuss the inverse effect
of TSOT, i.e., the generation of
heat current due to magnetization dynamics in ferromagnets with broken
inversion symmetry and SOI (Figure~1d). 
We refer to this effect as inverse thermal spin-orbit
torque (ITSOT).
While the SOT is given directly by the 
linear response of the torque to an applied electric field~\cite{ibcsoit},  
expressions for the ITSOT are more difficult to derive
because the energy current obtained from the Kubo formalism
contains also a ground-state contribution
that does not contribute to the heat current.
Analogous difficulties are known from the case of the
inverse anomalous Nernst effect, i.e., the generation of a 
heat current transverse to an applied electric field $\vn{E}$~\cite{ane_niu}. 
In this case the energy current obtained from the Kubo formalism 
contains besides the heat current also the material-dependent
part $-\vn{E}\times\vn{M}^{\rm orb}$ of the Poynting vector, 
where $\vn{M}^{\rm orb}$ is 
the orbital magnetization. This energy magnetization does not
contribute to the heat current and needs to be subtracted from
the Kubo linear 
response~\cite{ane_niu,thermoelectric_response_cooper,energy_magnetization_qin_niu_shi}.

When inversion symmetry is broken in magnets with
SOI the expansion of
the free energy $F$ in terms of the 
magnetization direction $\hat{\vn{n}}(\vn{r})$ 
and its
gradients contains a term linear in the gradients of magnetization,
the so-called Dzyaloshinskii-Moriya interation 
(DMI)~\cite{dmi_moriya,dmi_dzyalo}:
\bege\label{eq_first_order_free_energy}
F^{\rm DMI}(\vn{r})=
\sum_{j}
\vn{D}_{j}(\hat{\vn{n}}(\vn{r}))
\cdot
\left(
\hat{\vn{n}}(\vn{r})\times\frac{\partial \hat{\vn{n}}(\vn{r})}{\partial r_{j}}
\right),
\ee  
where $\vn{r}$ is the position and the index $j$ runs over the three
cartesian directions, i.e., $r_{1}=x, r_{2}=y, r_{3}=z$.
The DMI coefficients $\vn{D}_{j}$
can be expressed in terms of mixed 
Berry phases~\cite{mothedmisot,phase_space_berry}.
DMI does not only affect the magnetic structure by energetically
favoring spirals of a certain handedness but also enters
spin caloritronics effects~\cite{magnetization_pumping_dm_magnet,
magnon_mediated_dmi_torque}.
Here, we will show that DMI
gives rise to the ground-state energy current 
$\mathscr{J}^{\rm DMI}_{j}=-\vn{D}_{j}
\cdot
\left(
\hat{\vn{n}}\times\frac{\partial \hat{\vn{n}}}{\partial t}
\right)$
when 
magnetization precesses. This DMI
energy current
needs to be subtracted from the
linear response of the energy current in order to obtain the
ITSOT heat current.

This work is structured as follows.
In section~\ref{sec_ground_state_energy_current} we show that
magnetization dynamics drives a ground-state energy current 
associated with DMI and we highlight its formal similarities with the
material-dependent part of the Poynting vector.  
In section~\ref{sec_itsot} we develop the theory of ITSOT. We derive
the energy current based on the Kubo linear-response formalism
and subtract $\mathscrbf{J}^{\rm DMI}$ in order to extract the heat current.
In section~\ref{sec_time_dependent} we show that the expressions of
DMI and orbital magnetization can also be derived elegantly
by equating the energy currents obtained from linear response theory to
$\mathscrbf{J}^{\rm DMI}$ and
$-\vn{E}\times\vn{M}^{\rm orb}$, respectively. 
In section~\ref{section_ab_initio} we present \textit{ab-initio}
calculations of TSOT and ITSOT in Mn/W(001) magnetic bilayers.
\section{Ground-state energy current associated with the
  Dzyaloshinskii-Moriya interaction}
\label{sec_ground_state_energy_current}
To be concrete, we consider
a flat cycloidal spin spiral propagating along the $x$ direction.
The magnetization direction is given by
\bege\label{eq_spin_spiral_cycloid}
\hat{\vn{n}}_{\rm c}(\vn{r})=\hat{\vn{n}}_{\rm c}(x)=
\begin{pmatrix}
\sin(qx)\\
0\\
\cos(qx)
\end{pmatrix},\\
\ee
where $q$ is the spin-spiral wavenumber, i.e., the 
inverse wavelength of the spin spiral multiplied by $2\pi$.
The free energy contribution $F^{\rm DMI}(\vn{r})$ given 
in \eqref{eq_first_order_free_energy} simplifies
for the spin spiral of \eqref{eq_spin_spiral_cycloid} as follows:
\bege\label{eq_first_order_free_energy_cycloidal_along_x}
\begin{aligned}
F^{\rm DMI}(\vn{r})&=F^{\rm DMI}(x)=
\vn{D}_{x}(\hat{\vn{n}}_{\rm c}(x))
\!\cdot\!
\left[
\hat{\vn{n}}_{\rm c}(x)\!\times\!\frac{\partial \hat{\vn{n}}_{\rm c} (x)}{\partial x}
\right]=\\
&=q
\vn{D}_{x}(\hat{\vn{n}}_{\rm c} (x)) \cdot \hat{\vn{e}}_{y}
=q \mathscr{D}_{xy}(\hat{\vn{n}}_{\rm c} (x)),
\end{aligned}
\ee
where $\hat{\vn{e}}_{y}$ is the unit vector pointing in $y$ direction
and we defined
$\mathscr{D}_{ij}(\hat{\vn{n}})=\vn{D}_{i}(\hat{\vn{n}})\cdot\hat{\vn{e}}_{j}$.
Whether $\mathscr{D}_{xy}$ is nonzero or not depends on crystal symmetry.
The tensor $\mathscr{D}_{ij}(\hat{\vn{n}})$ is axial and of second rank like
the SOT torkance tensor~\cite{mothedmisot}. Additionally, it is
even under magnetization reversal, i.e.,
$\mathscr{D}_{ij}(\hat{\vn{n}})=\mathscr{D}_{ij}(-\hat{\vn{n}})$.
Therefore, $\mathscr{D}_{ij}(\hat{\vn{n}})$ has the same symmetry properties
as the even SOT torkance~\cite{ibcsoit}. According to 
\eqref{eq_first_order_free_energy_cycloidal_along_x}
the cycloidal spiral of \eqref{eq_spin_spiral_cycloid} is affected
by DMI if $\mathscr{D}_{xy}$ is nonzero. This is the case e.g.\ for magnetic
bilayers such as Mn/W(001) and Co/Pt(111) (the interface normal 
points in $z$ direction), where also the component
$t_{yx}$ of the even SOT torkance tensor is nonzero~\cite{ibcsoit,mothedmisot,
dmi_mnw_ferriani,dmi_copt_thiaville}.

\begin{figure}
\flushright
\includegraphics[width=\myfigurewidth,trim=4cm 0cm 1cm 0cm,clip]{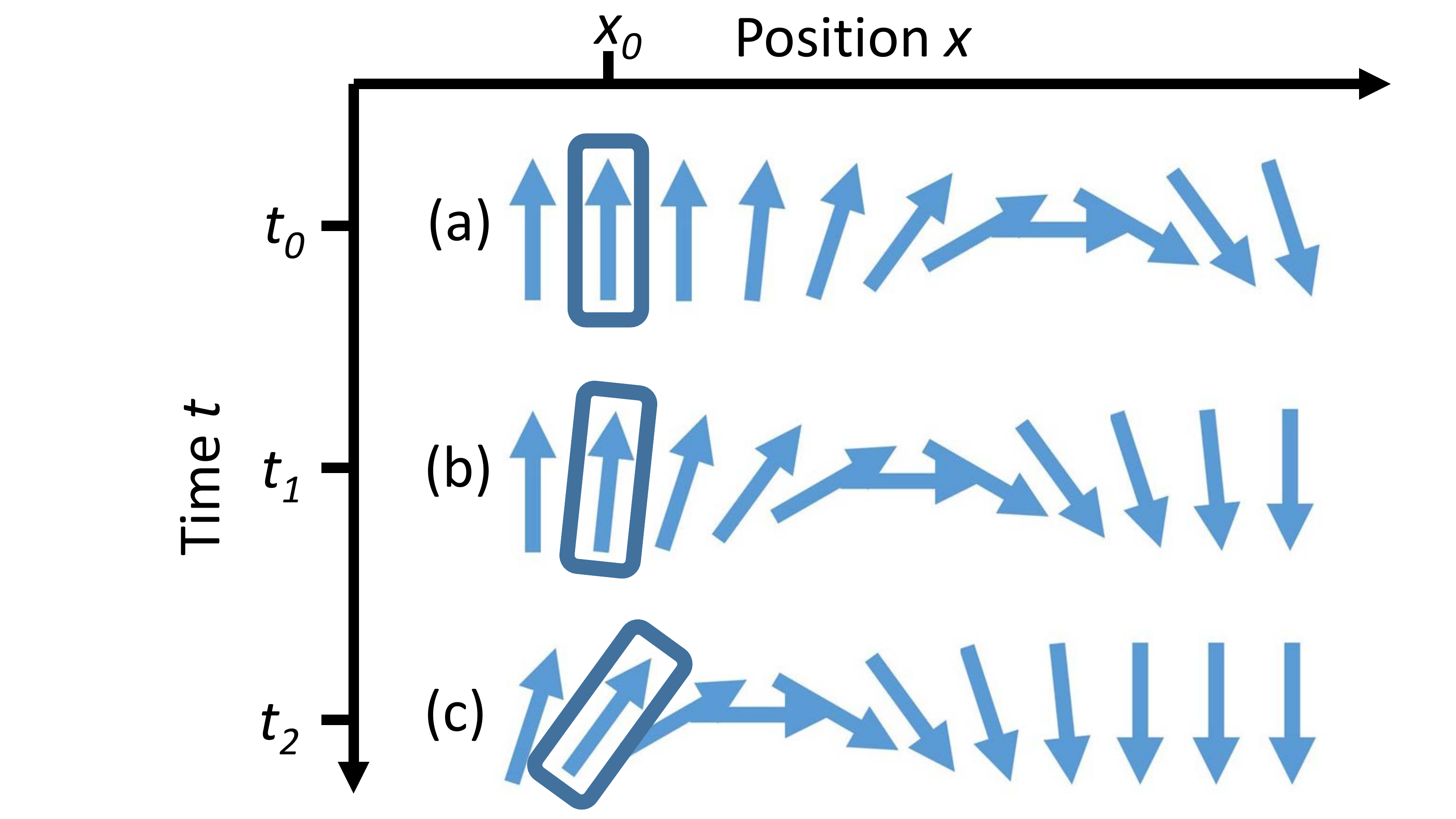}
\caption{\label{figure1}
Illustration of a Neel-type domain wall that moves
into the negative $x$ direction. Arrows represent the
magnetization direction $\hat{\vn{n}}(x,t)$.
$\hat{\vn{n}}(x_0,t)$ is highlighted by oval boxes.
(a) $\hat{\vn{n}}(x,t_0)=\hat{\vn{n}}_{0}(x)$ is locally
collinear at $x_0$ and therefore $F^{\rm DMI}(x_0,t_0)=0$.
(b) $\hat{\vn{n}}(x,t_1)=\hat{\vn{n}}_{0}(x-wt_1)$
starts to become noncollinear at $x_0$
and therefore $F^{\rm DMI}(x_0,t_1)\ne 0$.
(c) $\hat{\vn{n}}(x,t_2)$ is strongly noncollinear at $x_0$.
}
\end{figure}

We consider now a Neel-type domain
wall that moves with velocity $w<0$ in $x$ direction.
The magnetization direction at time $t_{0}=0$, which we denote by
$\hat{\vn{n}}_{0}(x)$, is
illustrated in Figure~\ref{figure1}a. $\hat{\vn{n}}_{0}(x)$ can be
interpreted as a modification of $\hat{\vn{n}}_{\rm c}(x)$ 
(\eqref{eq_spin_spiral_cycloid}),
where the $q$-vector depends on position:
\bege
\hat{\vn{n}}_{0}(x)=
\begin{pmatrix}
\sin(q(x)x)\\
0\\
\cos(q(x)x)\\
\end{pmatrix}.
\ee
Since the domain wall moves with velocity $w$, the magnetization direction
$\hat{\vn{n}}(x,t)$ at position $x$ and time $t$ is given by
\bege
\hat{\vn{n}}(x,t)=\hat{\vn{n}}_{0}(x-wt).
\ee
In Figure~\ref{figure1} we discuss the magnetization direction at position
$x_0$ at the three times $t_0=0$, $t_1>t_0$ and $t_2>t_1$. At time $t_0=0$ the
domain wall is far away from $x_0$. Therefore, the magnetization is collinear
at $x_0$ and $F^{\rm DMI}(x_0,t_0)=0$. At time $t_1$ the domain wall starts to
arrive at $x_0$. Consequently, the magnetization
gradient $\partial \hat{\vn{n}}(x_0,t_1)/\partial x_0$
becomes nonzero and thus $F^{\rm DMI}(x_0,t_1)\neq 0$. Due to the motion of
the domain wall the DMI contribution $F^{\rm DMI}(x,t)$ to the free energy
is time dependent:
How much DMI free energy
is stored at a given position in the magnetic structure is determined
by the local gradient of magnetization, which moves together
with the magnetic structure. 
The partial derivative of $F^{\rm DMI}(x,t)$ with respect to time is
given by
\begin{gather}\label{eq_dmi_current_continuity_x}
\begin{aligned}
&\frac{\partial F^{\rm DMI}(x,t)}{\partial t}
=
\\
&=\!
\vn{D}_{x}(\hat{\vn{n}}_0(x\!-\!wt))
\cdot
\left[
\hat{\vn{n}}_0(x\!-\!wt)
\!
\times
\!
\frac{\partial^2 \hat{\vn{n}}_0(x-wt)}{\partial x\partial t}
\right]+
\\&\quad+\!  
\frac{\partial\vn{D}_{x}(\hat{\vn{n}}_0(x\!-\!wt))}{\partial t}
\cdot
\left[
\hat{\vn{n}}_0(x\!-\!wt)
\!
\times
\!
\frac{\partial \hat{\vn{n}}_0(x-wt)}{\partial x}
\right]=\\
&=\!
\vn{D}_{x}(\hat{\vn{n}}_0(x\!-\!wt))
\cdot
\left[
\hat{\vn{n}}_0(x\!-\!wt)
\!
\times
\!
\frac{\partial^2 \hat{\vn{n}}_0(x-wt)}{\partial x\partial t}
\right]+\\
&\quad+\!
\frac{\partial\vn{D}_{x}(\hat{\vn{n}}_0(x\!-\!wt))}{\partial x}
\cdot
\left[
\hat{\vn{n}}_0(x\!-\!wt)
\!
\times
\!
\frac{\partial \hat{\vn{n}}_0(x-wt)}{\partial t}
\right]=\\
&=
\frac{\partial}{\partial x}
\left\{
\vn{D}_{x}(\hat{\vn{n}}_0(x-wt))
\cdot
\left[
\hat{\vn{n}}_0(x-wt)
\!\times\!
\frac{\partial \hat{\vn{n}}_0(x-wt)}{\partial t}
\right]
\right\}=\\
&=-
\frac{\partial
}{\partial x}\mathscr{J}_{x}^{\rm DMI},
\end{aligned}\raisetag{6\baselineskip}
\end{gather}
where $\mathscr{J}_{x}^{\rm DMI}$ in the last line is the $x$ component of the
DMI energy current density
\bege\label{eq_dmi_energy_current}
\begin{aligned}
\mathscrbf{J}^{\rm DMI}
=&-\sum_{ij}
\hat{\vn{e}}_{j}
\mathscr{D}_{ji}(\hat{\vn{n}})
\left[\hat{\vn{e}}_{i}
\cdot
\left(
\hat{\vn{n}}
\times
\frac{\partial \hat{\vn{n}}}{\partial t}
\right)
\right]\\
=&-\mathscrbf{D}(\hat{\vn{n}})
\left(
\hat{\vn{n}}
\times
\frac{\partial \hat{\vn{n}}}{\partial t}
\right).
\end{aligned}
\ee
By considering additionally spirals propagating in $y$ and $z$
direction we find that the general form of
\eqref{eq_dmi_current_continuity_x} is
the continuity equation 
\bege\label{eq_dmi_energy_current_continuity}
\frac{\partial F^{\rm DMI}}{\partial t}
+
\vn{\nabla}\cdot \mathscrbf{J}^{\rm DMI}
=0
\ee
of the DMI energy current $\mathscrbf{J}^{\rm DMI}$.
According to \eqref{eq_dmi_energy_current} 
and \eqref{eq_dmi_energy_current_continuity}
the energy current $\mathscrbf{J}^{\rm DMI}$ is driven by
magnetization dynamics and its sources and sinks signal
the respective decrease and increase of DMI energy density. 
When we compute the energy current driven by magnetization
dynamics in section~\ref{sec_itsot} we therefore need to be aware
that this energy current contains $\mathscrbf{J}^{\rm DMI}$ in
addition
to the ITSOT heat current that we wish to determine.
Thus, we need
to subtract $\mathscrbf{J}^{\rm DMI}$ from the
energy current in order to extract the ITSOT heat current.

\begin{figure}
\flushright
\includegraphics[width=\myfigurewidth,trim=0cm 0cm 6.2cm 0cm,clip]{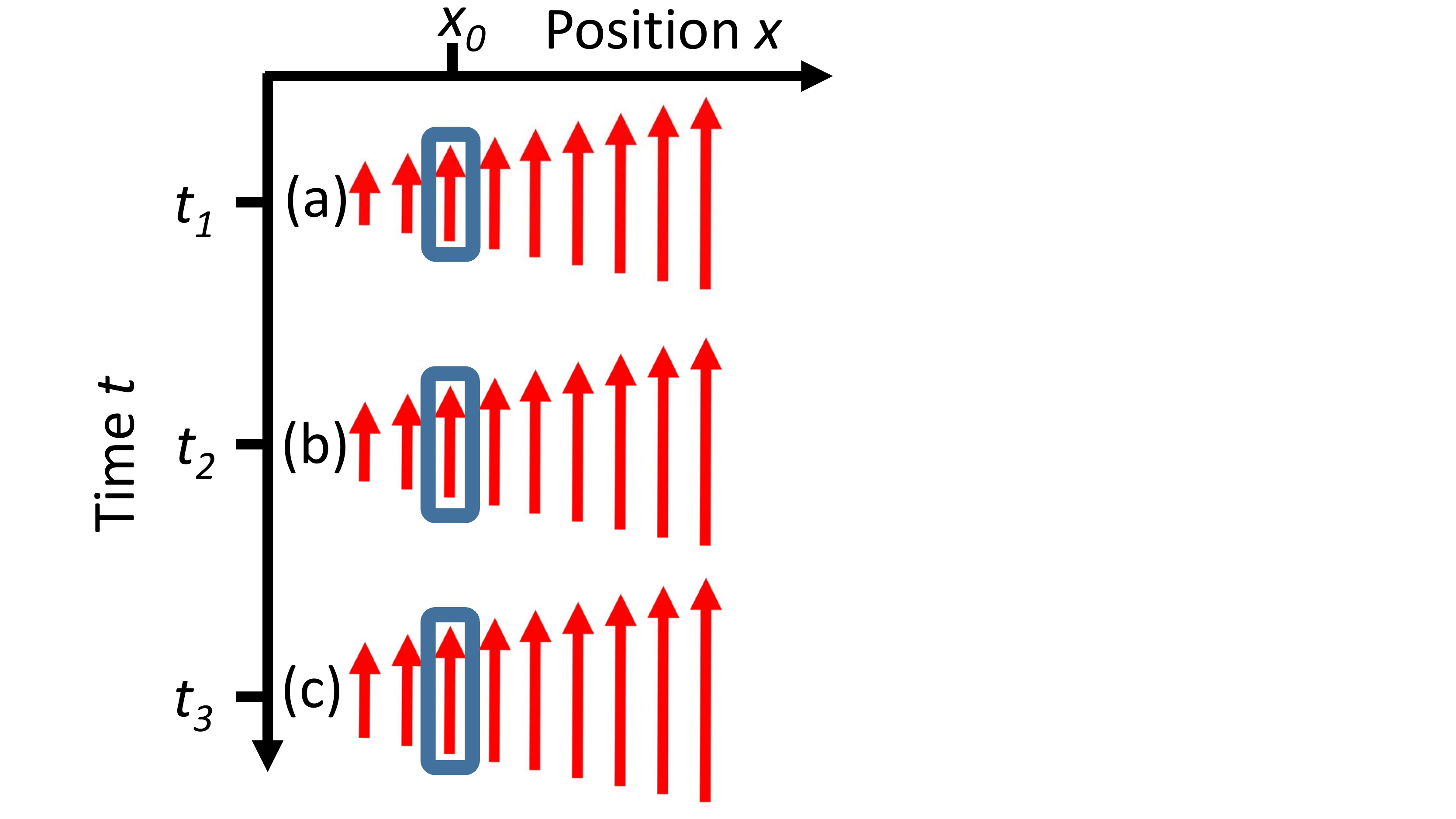}
\caption{\label{figure2}
Illustration of a magnetic field ramp that moves
into the negative $x$ direction.
Arrows represent the
magnetic field $B_{0}(x-wt)\hat{\vn{e}}_{z}$ at position $x$ and
time $t$. $B_{0}(x)$ describes
a ramp that increases linearly with $x$.
The magnetic field at position $x_0$ is highlighted by an oval box.
(a) Snapshot at time $t_1$.
(b) At time $t_2$ the magnetic field at position $x_0$ has increased
because the ramp has moved to the left since $t_1$. Consequently, also the energy
density $F_0^{\rm orb}(x_0-wt_2)$ is now different.
(c) At time $t_3$ the magnetic field at position $x_0$ has increased further.
}
\end{figure}

It is reassuring to verify that the material-dependent 
part $\mathscrbf{J}^{\rm orb}=-\vn{E}\times\vn{M}^{\rm orb}$ of the Poynting vector, which needs to
be subtracted from the energy current to obtain the heat current in
the case of the inverse anomalous Nernst effect~\cite{ane_niu}, can 
be identified
by arguments analogous to the above. We sketch this in the following.
The energy density due to the interaction between orbital
magnetization $\vn{M}^{\rm orb}$ and magnetic field $\vn{B}$ is given
by
\bege\label{eq_forb}
F^{\rm orb}(\vn{r},t)=-\vn{M}^{\rm orb}(\vn{r},t)\cdot \vn{B}(\vn{r},t).
\ee
We assume that the magnetic field is of the form
\bege
\vn{B}(\vn{r},t)=B_{0}(x-wt)\hat{\vn{e}}_{z},
\ee
i.e., the magnetic field at time $t$ can be obtained from the
magnetic field at time $t_0=0$ by shifting it by $wt$,
as illustrated in Figure~\ref{figure2}. 
Additionally, we assume that the orbital magnetization is of the
same form, i.e., $\vn{M}^{\rm orb}_{\phantom{0}}(\vn{r},t)=M^{\rm
  orb}_{0}(x-wt)\hat{\vn{e}}_{z}$.
Consequently, also $F^{\rm orb}_{\phantom{0}}(\vn{r},t)=F^{\rm
  orb}_{0}(x-wt)$.
$\vn{B}(\vn{r},t)$ can be
expressed as $\vn{B}(\vn{r},t)=\vn{\nabla}\times \vn{A}(\vn{r},t)$ in terms of the vector potential 
\bege
\vn{A}(\vn{r},t)=\hat{\vn{e}}_{y}\int_{0}^{x-wt} B_{0}(x')dx'.
\ee
Due to the motion of the profile of $\vn{B}(\vn{r},t)$ the energy
density in \eqref{eq_forb} changes as a function of time. The
partial derivative of $F^{\rm orb}_{\phantom{0}}(\vn{r},t)$ with
respect to time is
\bege
\begin{aligned}
\frac{\partial F^{\rm orb}}{\partial t}
&=-\frac{\partial\vn{M}^{\rm orb}}{\partial t}\cdot \vn{B}
-\vn{M}^{\rm orb}\cdot
\frac{\partial\vn{B}}{\partial t}\\
&=w\frac{\partial\vn{M}^{\rm orb}}{\partial x}\cdot [\vn{\nabla}\times\vn{A}]
+\vn{M}^{\rm orb}\cdot
[\vn{\nabla}\times\vn{E}]\\
&=w\frac{\partial\vn{M}^{\rm orb}}{\partial x}\cdot
\hat{\vn{e}}_{z}\frac{\partial A_y}{\partial x}
+\vn{M}^{\rm orb}\cdot
[\vn{\nabla}\times\vn{E}]\\
&=-\frac{\partial\vn{M}^{\rm orb}}{\partial x}\cdot
\hat{\vn{e}}_{z}\frac{\partial A_y}{\partial t}
+\vn{M}^{\rm orb}\cdot
[\vn{\nabla}\times\vn{E}]\\
&=-\vn{E}\cdot[\vn{\nabla}\times\vn{M}^{\rm orb}]
+\vn{M}^{\rm orb}\cdot
[\vn{\nabla}\times\vn{E}]\\
&=\vn{\nabla}\cdot[\vn{E}\times\vn{M}^{\rm orb}],
\end{aligned}
\ee
where we used the Maxwell equation
$\vn{\nabla}\times\vn{E}+\frac{\partial\vn{B}}{\partial t}=0$ and
$\vn{E}=-\frac{\partial \vn{A}}{\partial t}$ valid in Weyl's temporal
gauge
with scalar potential set to zero. Thus, 
\bege\label{eq_conti_orb}
\frac{\partial F^{\rm orb}}{\partial t}+\vn{\nabla}\cdot
\mathscrbf{J}^{\rm orb}=0
\ee
with 
\bege\label{eq_curr_orb}
\mathscrbf{J}^{\rm orb}=-\vn{E}\times\vn{M}^{\rm orb},
\ee
 as expected.

In the following we discuss several additional formal analogies and
similarities between
DMI, classical electrodynamics and orbital magnetization.
We introduce the tensors $\mathscrbf{C}(\vn{r})$ 
and $\bar{\mathscrbf{C}}(\vn{r})$
with
elements
\bege\label{eq_noco_tensor}
\mathscr{C}_{ij}(\vn{r})=
\hat{\vn{e}}_{i}
\cdot
\left[
\hat{\vn{n}}(\vn{r})
\times
\frac{\partial \hat{\vn{n}}(\vn{r})}{\partial r_{j}}
\right]
\ee
and
\bege
\bar{\mathscr{C}}_{ij}(\vn{r})=
\frac{\partial \hat{n}_{i}(\vn{r})}{\partial r_{j}}
\ee
to quantify the noncollinearity of $\hat{\vn{n}}(\vn{r})$.
$\mathscrbf{C}$ and $\bar{\mathscrbf{C}}$
are related through the matrix
\bege
\mathscrbf{K}(\hat{\vn{n}})=
\begin{pmatrix}
0 &-\hat{n}_3 &\hat{n}_2\\
\hat{n}_3 &0 &-\hat{n}_1\\
-\hat{n}_2 &\hat{n}_1 &0
\end{pmatrix}
\ee
as $\mathscrbf{C}=\mathscrbf{K}\bar{\mathscrbf{C}}$.
The free energy $F^{\rm DMI}(\vn{r})$ can be expressed
in terms of $\mathscrbf{C}$ and $\mathscrbf{D}$
as follows:
\bege\label{eq_dmi_free_energy_c_tensor}
\begin{aligned}
&F^{\rm DMI}(\vn{r})=\sum_{j}
\vn{D}_{j}(\vn{r})
\cdot
\left[
\hat{\vn{n}}(\vn{r})
\times
\frac{\partial \hat{\vn{n}}(\vn{r})}{\partial r_{j}}
\right]\\
&\quad=\sum_{ij}
\mathscr{D}_{ji}(\vn{r})\hat{\vn{e}}_{i}
\cdot
\left[
\hat{\vn{n}}(\vn{r})
\times
\frac{\partial \hat{\vn{n}}(\vn{r})}{\partial r_{j}}
\right]\\
&\quad=\sum_{ij}
\mathscr{D}_{ji}(\vn{r})
\mathscr{C}_{ij}(\vn{r})
=
{\rm Tr}[\mathscrbf{D}(\vn{r})\mathscrbf{C}(\vn{r})]=\\
&\quad=
{\rm Tr}[
\mathscrbf{D}(\vn{r})
\mathscrbf{K}(\hat{\vn{n}}(\vn{r}))
\bar{\mathscrbf{C}}(\vn{r})]
=
{\rm Tr}[
\bar{\mathscrbf{D}}(\vn{r})
\bar{\mathscrbf{C}}(\vn{r})],
\\
\end{aligned}
\ee
where we
defined $\bar{\mathscrbf{D}}=\mathscrbf{D}\mathscrbf{K}$.
Similarly, $\mathscrbf{J}^{\rm DMI}$ in \eqref{eq_dmi_energy_current}
can be expressed in terms of $\bar{\mathscrbf{D}}$ as
\bege\label{eq_dmi_energy_current_bar}
\begin{aligned}
\mathscrbf{J}^{\rm DMI}
=&-\mathscrbf{D}
\left(
\hat{\vn{n}}
\times
\frac{\partial \hat{\vn{n}}}{\partial t}
\right)
=
-\bar{\mathscrbf{D}}
\frac{\partial \hat{\vn{n}}}{\partial t}
.
\end{aligned}
\ee
The energy density $F^{\rm
  orb}=-\vn{M}^{\rm orb}\cdot(\vn{\nabla}\times\vn{A})$ in \eqref{eq_forb} 
involves the curl of the vector potential $\vn{A}$,
while the material-dependent part of the Poynting vector, i.e.,
$\mathscrbf{J}^{\rm orb}
\!=\!
-\vn{E}\!\times\!\vn{M}^{\rm orb}=\frac{\partial\vn{A}}{\partial t}\!\times\!\vn{M}^{\rm orb}$,
involves the time-derivative of $\vn{A}$. 
Similarly, the spatial derivatives $\partial\hat{\vn{n}}/\partial
r_{i}$ enter $F^{\rm DMI}$ in \eqref{eq_dmi_free_energy_c_tensor}
via the tensor $\bar{\mathscrbf{C}}$ while the temporal derivative $\partial\hat{\vn{n}}/\partial
t$ enters $\mathscrbf{J}^{\rm DMI}$ in \eqref{eq_dmi_energy_current_bar}.
Thus, in the theory of DMI the magnetization direction $\hat{\vn{n}}$
plays the role of an effective vector potential.

The curl of orbital magnetization constitutes a bound 
current $\vn{J}^{\rm mag}=\vn{\nabla}\times \vn{M}^{\rm orb}$ that
does not contribute to
electronic transport. Hence it needs to be subtracted from the linear
response electric current driven by gradients in temperature or chemical
potential
in order to obtain the measurable electric current~\cite{ane_niu}.
Similarly, the spatial derivatives $\Torque^{\rm bound}_{j}
=\sum_{i}\frac{\partial}{\partial r_{i}}\mathscr{D}_{ij}
=
\vn{\nabla}\cdot\left[
\mathscrbf{D}\hat{\vn{e}}_{j}
\right]
$ that result from the presence of
gradients in temperature or chemical potential constitute torques
that are not measurable and need to be subtracted from the
total linear response to temperature or chemical potential
gradients in order to obtain the measurable torque~\cite{mothedmisot}.
Table~\ref{table_compare_om_dmi} summarizes the formal analogies and
similarities
between the orbital magnetization and DMI.

\renewcommand{\arraystretch}{1.5}
\begin{table}
\caption{\label{table_compare_om_dmi} Formal analogies between the 
theories of orbital magnetization (OM) and
  Dzyaloshinskii-Moriya interaction (DMI). The vector potential
  $\vn{A}$ is assumed to satisfy Weyl's temporal gauge, hence the scalar
  potential is set to zero.}
\begin{indented}
\item[]\begin{tabular}{@{} lll }
\br 
&OM
& DMI
\\ \mr 
'vector potential'
&$\vn{A}$
&$\hat{\vn{n}}$
\\ \hline
'magnetic' field
&$\vn{B}=\vn{\nabla}\times\vn{A}$
&$\bar{\mathscr{C}}_{ij}=\frac{\partial \hat{n}_{i}}{\partial r_{j}}$
\\ \hline
energy density
&$F^{\rm orb}
\!=\!
-\vn{M}^{\rm orb}\!\cdot\!\vn{B}$
&$F^{\rm DMI}={\rm Tr}[\bar{\mathscrbf{D}}\bar{\mathscrbf{C}}]$
\\ \hline
'electric' field
&$\vn{E}=
-\frac{\partial\vn{A}}{\partial t}$
&$\frac{\partial \hat{\vn{n}}}{\partial t}$
\\ \hline
energy current
&$
\mathscrbf{J}^{\rm orb}
\!\!=\!\!
-\vn{M}^{\rm orb}
\!\!\times\!\!\frac{\partial \vn{A}}{\partial t}$
&$\mathscrbf{J}^{\rm DMI}
\!=\!
-
\bar{\mathscrbf{D}}\frac{\partial \hat{\vn{n}} }{\partial t}$
\\ \hline
bound property
& $\vn{J}^{\rm mag}=\vn{\nabla}\times \vn{M}^{\rm orb}$
& $\Torque^{\rm bound}_{j}
=
\vn{\nabla}\cdot\left[
\mathscrbf{D}\hat{\vn{e}}_{j}
\right]
$
\\ \br 
\end{tabular}
\end{indented}

\end{table}

\section{Inverse thermal spin-orbit torque (ITSOT)}
\label{sec_itsot}
In ferromagnets with broken inversion symmetry and SOI,
a gradient in temperature $T$
leads to a torque $\vn{\Torque}$ on
the magnetization, the so-called thermal
spin-orbit torque (TSOT)~\cite{mothedmisot,fept_guillaume}: 
\bege\label{eq_tsot_define_beta}
\vn{\Torque}=-\vn{\beta}\vn{\nabla} T.
\ee
The inverse thermal spin-orbit torque (ITSOT) consists 
in the generation of heat current by
magnetization dynamics
in ferromagnets with broken inversion symmetry and SOI. 
The effect of magnetization dynamics can be described by the
time-dependent
perturbation $\delta H$ to the Hamiltonian $H$~\cite{ibcsoit}
\bege
\delta H= \frac{\sin(\omega t)}{\omega}
\left[
\hat{\vn{n}}
\times
\frac{\partial\hat{\vn{n}}}{\partial t}
\right]\cdot\vn{\mathcal{T}},
\ee
where $\vht{\mathcal{T}}(\vn{r})=\vn{m}\times \hat{\vn{n}}\Bxc^{\rm
  xc}(\vn{r})$ is 
the torque operator. 
$\Bxc^{\rm xc}(\vn{r})=\frac{1}{2\mu_{\rm
    B}}\left[V^{\rm eff}_{\rm minority}(\vn{r})-V^{\rm eff}_{\rm
    majority}(\vn{r}) \right]$
is the exchange field, i.e., the
difference between the potentials of minority and
majority electrons. 
$\vn{m}=-\mu_{\rm B}\vht{\sigma}$ is the spin magnetic moment operator,
$\mu_{\rm B}$ is the
Bohr magneton  and 
$\vht{\sigma}=(\sigma_x,\sigma_y,\sigma_z)^{\rm T}$
is the vector of Pauli
spin matrices.
The energy current $\mathscrbf{J}^{E}$ driven by magnetization
dynamics is thus given by
\bege\label{eq_energy_curr_kubo_mag_dyn2}
\mathscrbf{J}^{E}=-\mathscrbf{B}(\hat{\vn{n}})
\left[
\hat{\vn{n}}
\times
\frac{\partial\hat{\vn{n}}}{\partial t}
\right],
\ee
where the tensor $\mathscrbf{B}$ with elements
\bege\label{eq_kubo_thermal_B}
\mathscr{B}_{ij}(\hat{\vn{n}})=
\lim_{\omega\to 0}\!
\frac{{\rm Im}
G_{\mathcal{J}^{E}_{i}\!\!,
\mathcal{T}_{j}^{\phantom{\alpha}}}^{\rm R}
\!(\hbar\omega,\hat{\vn{n}})}{\hbar\omega}
\ee
describes the Kubo linear response of the
energy current operator
\bege\label{eq_energy_curr_opera}
\vn{\mathcal{J}}^{E}=\frac{1}{2V}[(H-\mu)\vn{v}+\vn{v}(H-\mu)]
\ee
to magnetization dynamics.
$\mu$ is the chemical potential, $\vn{v}$ is the velocity operator and
the retarded energy-current torque correlation-function
is given by
\bege
G_{\mathcal{J}^{E}_{i}\!\!,
\mathcal{T}_{j}^{\phantom{\alpha}}}^{\rm R}
\!(\hbar\omega,\hat{\vn{n}})=
-i\int\limits_{0}^{\infty}dt e^{i\omega t}
\left\langle
[
\mathcal{J}^{E}_{i}(t),\mathcal{T}^{\phantom{\alpha}}_{j}(0)
]_{-}
\right\rangle.
\ee
In \eqref{eq_kubo_thermal_B} we take the limit
frequency $\omega\rightarrow 0$, which is justified when the 
frequency is small compared to the inverse lifetime of 
electronic states, which is satisfied for magnetic bilayers at room 
temperature and frequency $\omega/(2\pi)$ in the GHz range.  

Within the independent particle approximation
\eqref{eq_kubo_thermal_B} becomes
$
\mathscr{B}^{\phantom{I}}_{ij}
\!=
\mathscr{B}^{\rm I(a)}_{ij}
\!+
\mathscr{B}^{\rm I(b)}_{ij}
\!+
\mathscr{B}^{\rm II}_{ij}
$, with
\begin{gather}\label{eq_kubo_linear_response_pumped_energy_current}
\begin{aligned}
\mathscr{B}^{\rm I(a)\phantom{I}}_{ij}
&\!\!\!=\!\phantom{-}\frac{1}{h}\int_{-\infty}^{\infty}
\!\!\!d\mathcal{E}\frac{df(\mathcal{E})}{d\mathcal{E}}
\,{\rm Tr}
\left\langle
\mathcal{J}^{E}_{i}
G^{\rm R}(\mathcal{E})
\mathcal{T}_{j}
G^{\rm A}(\mathcal{E})
\right\rangle
\\
\mathscr{B}^{\rm I(b)\phantom{I}}_{ij}
&\!\!\!=\!-\frac{1}{h}\int_{-\infty}^{\infty}
\!\!\!d\mathcal{E}\frac{df(\mathcal{E})}{d\mathcal{E}}
\,{\rm Re}
\,{\rm Tr}
\left\langle
\mathcal{J}^{E}_{i}
G^{\rm R}(\mathcal{E})
\mathcal{T}_{j}
G^{\rm R}(\mathcal{E})
\right\rangle
\\
\mathscr{B}^{\rm II\phantom{(a)}}_{ij}
&\!\!\!=\!-
\frac{1}{h}\int_{-\infty}^{\infty}
d\mathcal{E}f(\mathcal{E})
\,{\rm Re}
\,{\rm Tr}
\left\langle
\mathcal{J}^{E}_{i}G^{\rm R}(\mathcal{E})\mathcal{T}_{j}
\frac{dG^{\rm R}(\mathcal{E})}{d\mathcal{E}}\right.\\
 &\quad\quad\quad\quad\quad\quad\quad\quad\,-\left.
\mathcal{J}^{E}_{i}\frac{dG^{\rm R}(\mathcal{E})}{d\mathcal{E}}
\mathcal{T}_{j}
G^{\rm R}(\mathcal{E})
\right\rangle,
\end{aligned}\raisetag{6.2\baselineskip}
\end{gather}
where $G^{\rm R}(\mathcal{E})$ and $G^{\rm A}(\mathcal{E})$ are the
retarded and advanced single-particle Green functions, respectively.
$f(\mathcal{E})$ is the Fermi function. $\mathscrbf{B}$ contains
scattering-independent intrinsic contributions and, in the presence of
disorder, additional disorder-driven contributions.  
The intrinsic Berry-curvature contribution is given by
\begin{gather}\label{eq_b_intrinsic}
\begin{aligned}
\mathscr{B}^{\rm int}_{ij}
\!&=\!\frac{2\hbar}{\mathcal{N}}
\!\sum_{\vn{k}n}
\!\!\sum_{m\neq n}
\!\!f_{\vn{k}n}
\text{Im}
\frac{
\langle \psi_{\vn{k}n}  |\mathcal{T}_{j}| \psi_{\vn{k}m}  \rangle
\langle \psi_{\vn{k}m}  |\mathcal{J}^{E}_{i}| \psi_{\vn{k}n}  \rangle
}
{
(\mathcal{E}_{\vn{k}m}-\mathcal{E}_{\vn{k}n})^2
}
\\
&=\frac{1}{\mathcal{N}V}
\!\!\sum_{\vn{k}n}
f_{\vn{k}n}
\left[
A_{\vn{k}nji}-(\mathcal{E}_{\vn{k}n}-\mu)B_{\vn{k}nji}
\right],
\end{aligned}\raisetag{2\baselineskip}
\end{gather}
where
\bege\label{eq_akn_kubo}
A_{\vn{k}nij}=\hbar\sum_{m\neq n}\text{Im}
\left[
\frac{
\langle \psi_{\vn{k}n}  |\mathcal{T}_{i}| \psi_{\vn{k}m}  \rangle
\langle \psi_{\vn{k}m}  |v_{j}| \psi_{\vn{k}n}  \rangle
}
{
\mathcal{E}_{\vn{k}m}-\mathcal{E}_{\vn{k}n}
}
\right]
\ee
and
\bege\label{eq_bkn_kubo}
B_{\vn{k}nij}
=-2\hbar\sum_{m\neq n}\text{Im}
\left[
\frac{
\langle \psi_{\vn{k}n}  |\mathcal{T}_{i}| \psi_{\vn{k}m}  \rangle
\langle \psi_{\vn{k}m}  |v_{j}| \psi_{\vn{k}n}  \rangle
}
{
(\mathcal{E}_{\vn{k}m}-\mathcal{E}_{\vn{k}n})^2
}
\right]
\ee
and $| \psi_{\vn{k}n}  \rangle$ are the Bloch wavefunctions 
with corresponding band energies $\mathcal{E}_{\vn{k}n}$,
$f_{\vn{k}n}=f(\mathcal{E}_{\vn{k}n})$, 
and $\mathcal{N}$ is the number of $\vn{k}$ points.

As discussed in section~\ref{sec_ground_state_energy_current} we
subtract $\mathscrbf{J}^{\rm DMI}$ (\eqref{eq_dmi_energy_current}) 
from $\mathscrbf{J}^{E}$ in order to obtain
the heat 
current $\mathscrbf{J}^{Q}$:
\bege\label{eq_subtract_dmi_for_heat}
\mathscrbf{J}^{Q}=\mathscrbf{J}^{E}-\mathscrbf{J}^{\rm DMI}
=-\tilde{\vn{\beta}}
\left[
\hat{\vn{n}}
\times
\frac{\partial\hat{\vn{n}}}{\partial t}
\right],
\ee
with
\bege\label{eq_beta_bmind}
\tilde{\vn{\beta}}=
\mathscrbf{B}-
\mathscrbf{D}.
\ee
Inserting the Berry-curvature expression of DMI~\cite{mothedmisot,phase_space_berry}
\bege\label{eq_dmi_finite_temperature}
\mathscr{D}_{ij}
\!=\!
\frac{1}{\mathcal{N}V}
\!\!\sum_{\vn{k}n}
\!\left\{
f_{\vn{k}n}A_{\vn{k}nji}
\!+\!\frac{1}{\beta}
\ln
\!\left[
1\!+\!e^{-\beta(\mathcal{E}_{\vn{k}n}-\mu)}
\right]
\!\!B_{\vn{k}nji}
\!\right\},
\ee
we obtain for the intrinsic contribution
\bege
\begin{aligned}\label{eq_thermo_beta_int}
\tilde{\beta}^{\rm int}_{ij}=&
\mathscr{B}^{\rm int}_{ij}-
\mathscr{D}^{\phantom{int}}_{ij}=\\
=&\frac{1}{\mathcal{N}V}
\sum_{\vn{k}n}
\bigg\{
f_{\vn{k}n}
\left[
A_{\vn{k}nji}-(\mathcal{E}_{\vn{k}n}-\mu)B_{\vn{k}nji}
\right]\\
-&
\!\left[
f_{\vn{k}n}A_{\vn{k}nji}
\!+\!\frac{1}{\beta}
\ln
\!\left[
1\!+\!e^{-\beta(\mathcal{E}_{\vn{k}n}-\mu)}
\right]
\!\!B_{\vn{k}nji}
\!\right]\bigg\}\\
=&-\frac{1}{\mathcal{N}V}
\sum_{\vn{k}n}
B_{\vn{k}nji}
\bigg\{
f_{\vn{k}n}
(\mathcal{E}_{\vn{k}n}-\mu)
+\\
&\quad\quad\quad\quad+\frac{1}{\beta}
\ln
\left[
1+e^{-\beta(\mathcal{E}_{\vn{k}n}-\mu)}
\right]
\bigg\},
\end{aligned}
\ee
where $\beta=(k_{\rm B}T)^{-1}$. Using
\begin{gather}
\begin{aligned}\label{eq_convert_log}
&f_{\vn{k}n}
(\mathcal{E}_{\vn{k}n}-\mu)+\frac{1}{\beta}
\ln
\left[
1+e^{-\beta(\mathcal{E}_{\vn{k}n}-\mu)}
\right]=\\
&=-\int_{-\infty}^{\mu}
d \mathcal{E}
f'(\mathcal{E}_{\vn{k}n}+\mu-\mathcal{E})
(\mathcal{E}_{\vn{k}n}-\mathcal{E})=\\
&=-\int_{-\infty}^{\mu}
\!\!\!\!d \mathcal{E}
\int_{-\infty}^{\infty}
\!\!\!\!d \mathcal{E}'
f'(\mathcal{E}'\!+\!\mu\!-\!\mathcal{E})
(\mathcal{E}'\!-\!\mathcal{E})
\delta(\mathcal{E}'\!-\!\mathcal{E}_{\vn{k}n})=\\
&=-
\int_{-\infty}^{\infty}
d \mathcal{E}'
f'(\mathcal{E}')
(\mathcal{E}'-\mu)
\Theta(\mathcal{E}'-\mathcal{E}_{\vn{k}n}),\\
\end{aligned}\raisetag{5\baselineskip}
\end{gather}
where
$\Theta$ is the Heaviside unit step function,
we can rewrite \eqref{eq_thermo_beta_int} as
\bege\label{eq_beta_itsot}
\tilde{\beta}_{ij}^{\rm int}(\hat{\vn{n}})=
-\frac{1}{eV}\int_{-\infty}^{\infty}
d\mathcal{E}
f'(\mathcal{E})
(\mathcal{E}-\mu)
t_{ji}^{\rm int}(\hat{\vn{n}},\mathcal{E}).
\ee
Here,
\bege
t_{ij}^{\rm int}(\hat{\vn{n}},\mathcal{E})
=-\frac{e}{\mathcal{N}}
\sum_{\vn{k}n}
\Theta(\mathcal{E}-\mathcal{E}_{\vn{k}n})
B_{\vn{k}nij}
\ee
is the intrinsic SOT torkance
tensor~\cite{ibcsoit,mothedmisot} at zero temperature as a function of
Fermi energy $\mathcal{E}$ and $e=|e|$ is the elementary positive charge.

The intrinsic TSOT and ITSOT are even in magnetization, i.e., 
$\tilde{\beta}_{ij}^{\rm int}(\hat{\vn{n}})=\tilde{\beta}_{ij}^{\rm int}(-\hat{\vn{n}})$.
\eqref{eq_kubo_linear_response_pumped_energy_current} contains an
additional contribution which is odd in 
magnetization, i.e., $\tilde{\beta}_{ij}^{\rm
  odd}(\hat{\vn{n}})=-\tilde{\beta}_{ij}^{\rm odd}(-\hat{\vn{n}})$,
and which is given by
\bege\label{eq_odd_itsot}
\tilde{\beta}^{\rm odd}_{ij}(\hat{\vn{n}})=\frac{1}{eV}\int_{-\infty}^{\infty}
d\mathcal{E}
f'(\mathcal{E})
(\mathcal{E}-\mu)
t_{ji}^{\rm odd}(\hat{\vn{n}},\mathcal{E}),
\ee
where $t_{ji}^{\rm odd}(\hat{\vn{n}},\mathcal{E})$ is the odd contribution to the
SOT torkance tensor as a function of Fermi energy~\cite{ibcsoit}.
The total $\tilde{\beta}_{ij}(\hat{\vn{n}})$ coefficient, i.e., the sum of all
contributions, is related to the total torkance
$t_{ji}(-\hat{\vn{n}},\mathcal{E})$
for magnetization in $-\hat{\vn{n}}$ direction
by
\bege\label{eq_total_itsot}
\tilde{\beta}_{ij}(\hat{\vn{n}})=-\frac{1}{eV}\int_{-\infty}^{\infty}
d\mathcal{E}
f'(\mathcal{E})
(\mathcal{E}-\mu)
t_{ji}(-\hat{\vn{n}},\mathcal{E}),
\ee 
which contains
\eqref{eq_beta_itsot} and 
\eqref{eq_odd_itsot} as special cases.

It is instructive to verify that the ITSOT described by 
\eqref{eq_total_itsot} 
is the Onsager-reciprocal of the TSOT
(\eqref{eq_tsot_define_beta}), where~\cite{mothedmisot}
\bege\label{eq_beta_tsot}
\beta_{ij}(\hat{\vn{n}})=
\frac{1}{e}\int_{-\infty}^{\infty}
d\mathcal{E}
f'(\mathcal{E})
\frac{(\mathcal{E}-\mu)}{T}
t_{ij}(\hat{\vn{n}},\mathcal{E}).
\ee
Comparison of \eqref{eq_total_itsot} and \eqref{eq_beta_tsot} yields
\bege\label{eq_betatilde_from_beta}
\vn{\beta}(\hat{\vn{n}})=-\frac{V}{T}[\tilde{\vn{\beta}}(-\hat{\vn{n}})]^{\rm T}
\ee
and thus
\bege
\begin{pmatrix}
-\mathscrbf{J}^{Q}\\
\vn{\Torque}/V
\end{pmatrix}
=
\begin{pmatrix}
T\vn{\lambda}(\hat{\vn{n}}) & \tilde{\vn{\beta}}(\hat{\vn{n}})\\
[\tilde{\vn{\beta}}(-\hat{\vn{n}})]^{\rm T} & -\vn{\Lambda}(\hat{\vn{n}})\\
\end{pmatrix}
\begin{pmatrix}
\frac{\vn{\nabla}T}{T}\\
\hat{\vn{n}}\times\frac{\partial\hat{\vn{n}}}{\partial t}
\end{pmatrix},
\ee
where $\vn{\lambda}$ is the thermal conductivity tensor
and $\vn{\Lambda}$ describes Gilbert damping and gyromagnetic
ratio~\cite{invsot}.
As expected, the response matrix
\bege
\mathscrbf{A}(\hat{\vn{n}})=
\begin{pmatrix}
T\vn{\lambda}(\hat{\vn{n}}) & \tilde{\vn{\beta}}(\hat{\vn{n}})\\
[\tilde{\vn{\beta}}(-\hat{\vn{n}})]^{\rm T} & -\vn{\Lambda}(\hat{\vn{n}})\\
\end{pmatrix}
\ee
satisfies the Onsager symmetry $\mathscrbf{A}(\hat{\vn{n}})=
[\mathscrbf{A}(-\hat{\vn{n}})]^{\rm T}$.

\eqref{eq_total_itsot} 
and \eqref{eq_subtract_dmi_for_heat} are the central result of
this section. Together, these two equations provide the recipe to
compute the heat current $\mathscrbf{J}^{\rm Q}$ driven by
magnetization dynamics $\partial\hat{\vn{n}}/\partial t$.
We discuss applications in section~\ref{section_ab_initio}.
\section{Using the ground-state energy currents to derive
expressions for DMI and orbital magnetization}
\label{sec_time_dependent}
The expression \eqref{eq_dmi_finite_temperature} for the DMI-spiralization tensor
$\mathscrbf{D}$ was derived both from semiclassics~\cite{phase_space_berry}
and static quantum mechanical perturbation theory~\cite{mothedmisot}.
Alternatively, the $T=0$ expression of
$\mathscrbf{D}$ can also be obtained elegantly and easily 
by invoking the third law of thermodynamics: For $T\rightarrow 0$ the ITSOT
must vanish, $\tilde{\vn{\beta}}\rightarrow 0$, because otherwise we
could pump heat at zero temperature and thereby violate Nernst's theorem.
Hence, $\mathscrbf{D}\rightarrow \mathscrbf{B}$ according 
to \eqref{eq_beta_bmind}. In other words, at $T=0$ the energy current
density $\mathscrbf{J}^{E}$ in
\eqref{eq_energy_curr_kubo_mag_dyn2}
is identical to the 
DMI energy current density $\mathscrbf{J}^{\rm DMI}=-
\mathscrbf{D}
\left(
\hat{\vn{n}}
\times
\frac{\partial \hat{\vn{n}}}{\partial t}
\right)$ 
because the heat current is zero.
Thus, at $T=0$ we obtain from \eqref{eq_b_intrinsic}
\bege
\mathscr{D}_{ij}=\mathscr{B}^{\rm int}_{ij}=\frac{1}{\mathcal{N}V}
\!\!\sum_{\vn{k}n}
f_{\vn{k}n}
\left[
A_{\vn{k}nji}-(\mathcal{E}_{\vn{k}n}-\mu)B_{\vn{k}nji}
\right],
\ee
which agrees with \eqref{eq_dmi_finite_temperature} at $T=0$.

Similarly, we can derive the $T=0$ expression of orbital magnetization
from the 
energy current $\mathscrbf{J}^{\rm orb}=-\vn{E}\times\vn{M}^{\rm orb}$
discussed in
\eqref{eq_curr_orb}: For $T\rightarrow 0$ the 
inverse anomalous Nernst effect (i.e., the generation of a transverse
heat current by an applied electric field) 
has to vanish according to the third
law of thermodynamics.
Hence, the energy current driven by an applied electric field at $T=0$
does not contain any heat current and is therefore 
identical to $\mathscrbf{J}^{\rm orb}$. 
We introduce the tensor $\mathscrbf{R}$ to describe the linear response of the
energy current $\mathscrbf{J}$ to an applied electric field $\vn{E}$, i.e.,
$\mathscrbf{J}=\mathscrbf{R}\vn{E}$.
We describe the effect of the electric field by the vector
potential $\vn{A}=-\vn{E}\sin(\omega t)/\omega$ and take the limit
$\omega\rightarrow 0$ later. The Hamiltonian density describing
the interaction between electric current density $\vn{J}$ and
vector potential is $-\vn{J}\cdot \vn{A}$, from which
we obtain the
time-dependent perturbation
\bege
\delta H=-\frac{\sin(\omega t)}{\omega}e\vn{E}\cdot\vn{v}.
\ee
Introducing the retarded energy-current velocity
correlation-function
\bege
G_{\mathcal{J}^{E}_{i}\!\!,v_{j}^{\phantom{\alpha}}}^{\rm R}
(\hbar\omega)
=
-i\int\limits_{0}^{\infty}dt e^{i\omega t}
\left\langle
[
\mathcal{J}^{E}_{i}(t),v^{\phantom{\alpha}}_{j}(0)
]_{-}
\right\rangle
\ee
we can write the elements of the tensor $\mathscrbf{R}$ as
\bege
\mathscr{R}_{ij}=
e
\lim_{\omega\to 0}\!
\frac{{\rm Im}
G_{\mathcal{J}^{E}_{i}\!\!,
v_{j}^{\phantom{\alpha}}}^{\rm R}
\!(\hbar\omega)}{\hbar\omega}.
\ee
This allows us to
determine $\mathscrbf{J}^{\rm orb}$ 
as $\mathscrbf{J}^{\rm orb}=\mathscrbf{R}^{\rm int}\vn{E}$,
where
the intrinsic Berry-curvature contribution to 
the response tensor $\mathscrbf{R}$ is given by
\bege
\begin{aligned}
\mathscr{R}_{ij}^{\rm int}=&
-\frac{2e\hbar}{\mathcal{N}}
\!\sum_{\vn{k}n}
f_{\vn{k}n}\!\!
\sum_{m\ne n}
\!\text{Im}
\frac{
\langle u_{\vn{k}n}  |
\mathcal{J}^{E}_{i}
| u_{\vn{k}m}  \rangle
\langle u_{\vn{k}m}  |
v_{j}^{\phantom{i}}
| u_{\vn{k}n}  \rangle
}
{
(\mathcal{E}_{\vn{k}m}-\mathcal{E}_{\vn{k}n})^2
}\\
=&
\frac{1}{\mathcal{N}V}\sum_{\vn{k}n}f_{\vn{k}n}
\left[
\mathscr{M}_{\vn{k}nij}
-
(\mathcal{E}_{\vn{k}n}-\mu)
\mathscr{N}_{\vn{k}nij}
\right]
,
\end{aligned}
\ee
with
\bege\label{eq_om_m_kn_kubo}
\mathscr{M}_{\vn{k}nij}=e\hbar\sum_{m\neq n}\text{Im}
\frac{
\langle u_{\vn{k}n}  |v_{i}| u_{\vn{k}m}  \rangle
\langle u_{\vn{k}m}  |v_{j}| u_{\vn{k}n}  \rangle
}
{
\mathcal{E}_{\vn{k}n}-\mathcal{E}_{\vn{k}m}
}
\ee
and
\bege\label{eq_om_n_kn_kubo}
\mathscr{N}_{\vn{k}nij}
=2e\hbar\sum_{m\neq n}\text{Im}
\frac{
\langle u_{\vn{k}n}  |v_{i}| u_{\vn{k}m}  \rangle
\langle u_{\vn{k}m}  |v_{j}| u_{\vn{k}n}  \rangle
}
{
(\mathcal{E}_{\vn{k}m}-\mathcal{E}_{\vn{k}n})^2
}.
\ee
From $\vn{M}^{\rm orb}\times\vn{E}=\mathscrbf{R}^{\rm int}\vn{E}$ we obtain
\bege\label{eq_orb_mag_tdpt}
\vn{M}^{\rm orb}=
-\frac{1}{2}
\hat{\vn{e}}_{k}^{\phantom{i}}
\epsilon^{\phantom{ij}}_{kij}
\mathscr{R}^{\rm int}_{ij}.
\ee
It is straightforward to verify that $\vn{M}^{\rm orb}$ given 
by \eqref{eq_orb_mag_tdpt} agrees to the $T=0$ expressions 
for orbital magnetization derived from quantum mechanical perturbation
theory~\cite{shi_quantum_theory_orbital_mag}, 
from semiclassics~\cite{ane_niu}, and within the 
Wannier representation~\cite{om_insulators_mlwfs,om_crystals_mlwfs}.

Combining the third law of thermodynamics with the
continuity equations \eqref{eq_dmi_energy_current_continuity} 
and \eqref{eq_conti_orb} provides thus an elegant way to derive
expressions for $\mathscrbf{D}$ and $\vn{M}^{\rm orb}$ at $T=0$.
We can extend these derivations to $T>0$ if we postulate that
the linear response to thermal gradients is described 
by Mott-like expressions.
In the case of the TSOT this Mott-like expression 
is \eqref{eq_beta_tsot}, while it 
is~\cite{ane_niu,ane_weischenberg,thermogalvanomagnetics_ebert}
\bege\label{eq_alpha_ane}
\alpha_{xy}=\frac{1}{e}
\int_{-\infty}^{\infty}
d\mathcal{E}
f'(\mathcal{E})
\frac{\mathcal{E}-\mu}{T}
\sigma_{xy}(\mathcal{E})
\ee
in the case of the anomalous Nernst effect, where
$\sigma_{xy}(\mathcal{E})$ is the zero-temperature
anomalous Hall conductivity as a function of Fermi
energy $\mathcal{E}$ and the anomalous Nernst current
due to a temperature gradient in $y$ direction
is $j_x=-\alpha_{xy}\partial T/\partial y$.
While
\eqref{eq_beta_tsot} and \eqref{eq_alpha_ane}
were, respectively, derived 
in the previous section and in  \cite{ane_niu}, 
we now instead consider it an axiom that
within the range of validity of the independent particle approximation
the linear response to thermal gradients is always described by 
Mott-like expressions. Thereby, the derivation in the present
section becomes independent from the derivation in the preceding section.  
Applying the Onsager reciprocity principle 
to \eqref{eq_beta_tsot} and \eqref{eq_alpha_ane} we find that the
ITSOT and the inverse anomalous Nernst effect
are, respectively, described by \eqref{eq_total_itsot}
and by
\bege\label{eq_iane_heat_curr}
\mathscr{J}^{Q}_{y}=T
\alpha^{\phantom{y}}_{xy}
E^{\phantom{y}}_{x}.
\ee
Employing the general identity \eqref{eq_convert_log}
(but in contrast
to section~\ref{sec_itsot} we now use it backwards)
we obtain 
\bege\label{eq_beta_tilde_intrinsic}
\begin{aligned}
\tilde{\beta}^{\rm int}_{ij}=&-\frac{1}{\mathcal{N}V}
\sum_{\vn{k}n}
B_{\vn{k}nji}
\bigg\{
f_{\vn{k}n}
(\mathcal{E}_{\vn{k}n}-\mu)
+\\
&\quad\quad\quad\quad+\frac{1}{\beta}
\ln
\left[
1+e^{-\beta(\mathcal{E}_{\vn{k}n}-\mu)}
\right]
\bigg\}
\end{aligned}
\ee 
from \eqref{eq_beta_itsot}
and, similarly, \eqref{eq_iane_heat_curr}
can be written as
\bege\label{eq_iane_heat_curr_curvature}
\begin{aligned}
\mathscr{J}^{Q}_{y}&=
-\frac{1}{\mathcal{N}V}
\sum_{\vn{k}n}
\mathscr{N}_{\vn{k}nyx}
\bigg\{
f_{\vn{k}n}
(\mathcal{E}_{\vn{k}n}-\mu)
+\\
&\quad\quad\quad\quad+\frac{1}{\beta}
\ln
\left[
1+e^{-\beta(\mathcal{E}_{\vn{k}n}-\mu)}
\right]
\bigg\}E_{x}^{\phantom{y}}.
\end{aligned}
\ee
 The finite-$T$ expressions of $\mathscrbf{D}$
and $\vn{M}^{\rm orb}$ are now easily obtained, respectively,
by subtracting the ITSOT heat current given by \eqref{eq_beta_tilde_intrinsic}
from the energy current \eqref{eq_b_intrinsic} and by subtracting
the heat current
\eqref{eq_iane_heat_curr_curvature}
from $\mathscr{J}^{\phantom{x}}_{y}=\mathscr{R}^{\rm int}_{yx}E^{\phantom{x}}_x$. This leads to \eqref{eq_dmi_finite_temperature} for the DMI
spiralization tensor
and to
\bege\label{eq_orb_mag_finite_t}
M_{z}^{\rm orb}
\!
=
\!
\frac{1}{\mathcal{N}V}
\!
\sum_{\vn{k}n}
\!
f_{\vn{k}n}
\!
\left\{
\!
\mathscr{M}_{\vn{k}nyx}
\!
+
\!
\frac{1}{\beta}
\mathscr{N}_{\vn{k}nyx}
\ln
\!
\left[
1
\!
+
\!
e^{-\beta(\mathcal{E}_{\vn{k}n}-\mu)}
\right]
\!
\right\}
\ee
for the orbital magnetization. \eqref{eq_orb_mag_finite_t}
agrees to the finite-$T$ expressions of $M_{z}^{\rm orb}$
derived elsewhere~\cite{shi_quantum_theory_orbital_mag,ane_niu}.
\section{Ab-initio calculations}
\label{section_ab_initio}
We investigate TSOT and ITSOT in a Mn/W(001) 
magnetic bilayer composed of one
monolayer of Mn deposited on 9 layers of W(001). 
The ground state of 
this system is magnetically noncollinear and can be
described by the cycloidal spin 
spiral \eqref{eq_spin_spiral_cycloid}~\cite{dmi_mnw_ferriani}. 
Based on phenomenological 
grounds~\cite{symmetry_considerations_PhysRevB.86.094406,
spin_motive_force_hals_brataas} 
we can expand torkance as well as
TSOT and ITSOT coefficients locally at a given point in space
in terms of $\hat{\vn{n}}$ 
and $\bar{\mathscrbf{C}}$:
\bege
\begin{aligned}
t^{\phantom{i}}_{ij}(\hat{\vn{n}},\bar{\mathscrbf{C}})&
=\sum_{k}t_{ijk}^{(1,0)}
\hat{n}_{k}^{\phantom{i}}
+
\sum_{kl}t_{ijkl}^{(0,1)}\bar{\mathscr{C}}_{kl}^{\phantom{i}}
+\\
&+\sum_{klm}t_{ijklm}^{(1,1)}\hat{n}^{\phantom{i}}_{k}
\bar{\mathscr{C}}^{\phantom{i}}_{lm}
+\sum_{k}t_{ijkl}^{(2,0)}\hat{n}^{\phantom{i}}_{k}\hat{n}^{\phantom{i}}_{l}
+\cdots.
\end{aligned}
\ee
The coefficients $t_{ijk}^{(1,0)}$, $t_{ijkl}^{(0,1)}$, 
$t_{ijklm}^{(1,1)}$,\dots in this expansion
can be extracted from magnetically collinear calculations.
Analogous expansions of the TSOT and ITSOT coefficients
are of the same form.
Here, we consider only $t_{ijk}^{(1,0)}$ and $t_{ijkl}^{(2,0)}$, which 
give rise to the following contribution to the torque $\vn{\tau}$:
\bege
\begin{aligned}
\vn{\tau}&=
t_{xx}^{\rm odd}(\hat{\vn{e}}_{z})
\hat{\vn{n}}\times
(\vn{E}\times \hat{\vn{e}}_{z})+\\
&+
t_{yx}^{\rm even}(\hat{\vn{e}}_{z})
\hat{\vn{n}}
\times
[
\hat{\vn{n}}\times
(\vn{E}\times \hat{\vn{e}}_{z})
],
\end{aligned}
\ee
where we used that for 
magnetization direction $\hat{\vn{n}}$ along $z$
it follows from symmetry considerations that
$t^{\phantom{x}}_{xx}=t^{\phantom{x}}_{yy}$,
$t^{\phantom{x}}_{xy}=-t^{\phantom{x}}_{yx}$,
$t^{\rm even}_{xx}=0$ and
$t^{\rm odd}_{yx}=0$.
The SOT in this system has already been discussed by us~\cite{ibcsoit}.
In order to obtain TSOT and ITSOT,
we calculate the torkance 
for the magnetically collinear ferromagnetic state with
magnetization direction $\hat{\vn{n}}$ set along $z$
as a function of Fermi energy and
use \eqref{eq_beta_tsot} and \eqref{eq_total_itsot} to
determine
the TSOT and ITSOT coefficients $\vn{\beta}$ 
and $\tilde{\vn{\beta}}$, 
respectively. 
Computational details of the density-functional
theory calculation of the electronic structure as well as technical
details of the torkance calculation
are given in  \cite{ibcsoit}. 
The torkance calculation is performed with the
help of Wannier functions~\cite{wannier90,WannierPaper}
and a quasiparticle broadening of $\Gamma=25$~meV is applied.

\begin{figure}
\flushright
\includegraphics[width=\myfigurewidth,trim=0cm 0cm 7cm 18cm,clip]{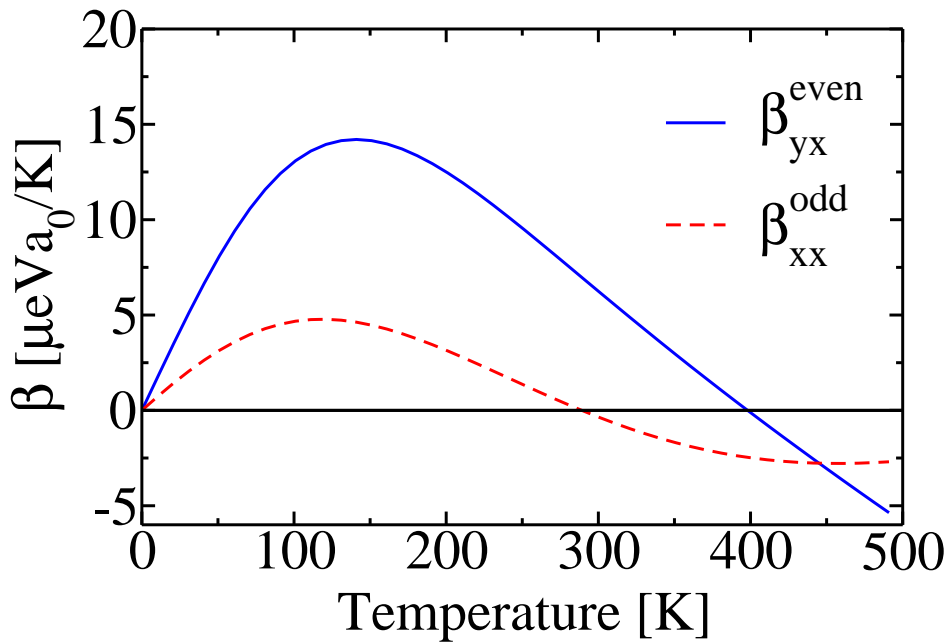}
\caption{\label{figuretsot} Thermal torkance $\beta$ vs.\ temperature 
of a Mn/W(001) magnetic bilayer for
magnetization in $z$ direction. 
Solid line: Even component $\beta_{yx}^{\rm even}$ of the
thermal torkance.
Dashed line: Odd component $\beta_{xx}^{\rm odd}$ of the
thermal torkance.
$\beta$ is plotted in units of
$\mu eVa_{0}/K=8.478\times 10^{-36}$Jm/K, where $a_{0}$ is Bohr's radius.
}
\end{figure}

Due to symmetry it suffices to discuss the
TSOT coefficients $\beta_{yx}^{\rm even}$ 
and $\beta_{xx}^{\rm odd}$, which are shown in Figure~\ref{figuretsot} as a 
function of temperature.
For small temperatures we find $\beta_{ij}\propto T$ as expected from
\bege\label{eq_low_temp_beta}
\beta_{ij}\simeq-\frac{\pi^2 k^2_{\rm B}T}{3e}
\frac{\partial\, t_{ij}}
{\partial\,\mu},
\ee
which is obtained from \eqref{eq_beta_tsot}
using the Sommerfeld expansion.
Slightly above 100K both 
$\beta_{yx}^{\rm even}$ and $\beta_{xx}^{\rm odd}$
stop following the linear behavior of the low temperature
expansion \eqref{eq_low_temp_beta}: After reaching
a maximum both 
$\beta_{yx}^{\rm even}$ and $\beta_{xx}^{\rm odd}$ 
decrease and finally change sign.
At $T=300$K the thermal torkances are
$\beta_{yx}^{\rm even}=5.24\times 10^{-35}$Jm/K
and
$\beta_{xx}^{\rm odd}=-3.21\times 10^{-36}$Jm/K.
Thermal torkances of comparable magnitude have been
determined in calculations on FePt/Pt 
magnetic bilayers~\cite{fept_guillaume}.

Using \eqref{eq_betatilde_from_beta} and 
the volume of the unit cell of $V=1.58\times 10^{-28}$m$^3$ 
to convert the TSOT coefficients into ITSOT coefficients,
we obtain
$\tilde{\beta}_{yx}^{\rm even}=-99.49\mu$J/m$^2$
and $\tilde{\beta}_{xx}^{\rm odd}=-6.09\mu$J/m$^2$
at $T=300$K.
When the magnetization precesses 
around the $z$ axis
in ferromagnetic resonance (this situation is sketched in Figure~1d)
with frequency $\omega$
and cone angle $\theta$
according to
\bege
\hat{\vn{n}}(t)=
[
\sin(\theta)\cos(\omega t),
\sin(\theta)\sin(\omega t),
\cos(\theta)
]^{\rm T},
\ee
the following ITSOT heat current
is obtained from \eqref{eq_subtract_dmi_for_heat} in the limit of
small $\theta$:
\bege\label{eq_fmr_heat_curr_z}
\begin{aligned}
\mathscr{J}_{x}^{Q}&=
\omega
\theta
\left[
\tilde{\beta}_{xx}^{\rm odd}\cos(\omega t)
-
\tilde{\beta}_{yx}^{\rm even}\sin(\omega t)
\right]
\\
\mathscr{J}_{y}^{Q}&=
\omega
\theta
\left[
\tilde{\beta}_{yx}^{\rm even}\cos(\omega t)
+
\tilde{\beta}^{\rm odd}_{xx}\sin(\omega t)
\right],
\end{aligned}
\ee
where we made use 
of $\tilde{\beta}^{\phantom{\rm od}}_{xx}=
\tilde{\beta}^{\phantom{\rm od}}_{yy}=\tilde{\beta}^{\rm odd}_{xx}$
and
$-\tilde{\beta}^{\phantom{\rm ev}}_{xy}=
\tilde{\beta}^{\phantom{\rm ev}}_{yx}=\tilde{\beta}^{\rm even}_{yx}$,
which follows from symmetry considerations.
Using the ITSOT coefficients $\tilde{\beta}^{\rm even}_{yx}$ 
and $\tilde{\beta}^{\rm odd}_{xx}$ 
determined above at $T=300$K we can determine the 
amplitudes of $\mathscr{J}_{x}^{Q}$ and $\mathscr{J}_{y}^{Q}$.
Assuming a cone angle of $1^{\circ}$ and a 
frequency of $\omega=2\pi\cdot$5GHz
we find that the amplitude of the oscillating heat 
current density $\mathscr{J}_{x}^{Q}$ is
\bege
\omega\theta
\sqrt{
\left(
\tilde{\beta}^{\rm even}_{yx}
\right)^2
+
\left(
\tilde{\beta}^{\rm odd}_{xx}
\right)^2
}\approx 55\frac{\rm kW}{{\rm m}^2}.
\ee
The heat current density $\mathscr{J}_{y}^{Q}$ has the same amplitude.
We can use the thermal conductivity of 
bulk W of $\lambda_{xx}$=174~W/(Km)~\cite{ho_powell_liley} at $T$=300~K
to estimate the temperature gradient needed to drive a heat
current of this magnitude: (55kW/m$^2$)/$\lambda_{xx}$=316~K/m.
The thickness of the Mn/W(001) film is 1.58~nm.
The amplitude of the heat current per length 
flowing in $x$ direction is thus
55~kW/m$^2\cdot$1.58~nm$\approx 87\mu$W/m.
These estimates suggest that $\mathscr{J}^{Q}_{\phantom{y}}$ is
measurable in ferromagnetic resonance experiments.

According to \eqref{eq_fmr_heat_curr_z} the heat current can be
made larger by increasing the cone angle.
However, in ferromagnetic resonance experiments the cone angle $\theta$ is 
small.
Therefore, we estimate the heat current driven by a flat
cycloidal spin spiral that moves with velocity $w$ in $x$ direction.
Its magnetization direction is given by
\bege
\hat{\vn{n}}_{\rm c}(\vn{r},t)=\hat{\vn{n}}_{\rm c}(x,t)=
\begin{pmatrix}
\sin(qx-wt)\\
0\\
\cos(qx-wt)
\end{pmatrix}.\\
\ee
With $\hat{\vn{n}}_{\rm c}(\vn{r},t)\times 
\partial \hat{\vn{n}}_{\rm c}(\vn{r},t)/\partial t=wq\hat{\vn{e}}_{y}$
we get
\bege
\mathscr{J}^{\rm Q}_{x}=-\tilde{\beta}^{\rm even}_{xy}wq
\ee
from \eqref{eq_subtract_dmi_for_heat},
i.e., a constant-in-time heat current in $x$ direction. 
Using $\tilde{\beta}_{xy}^{\rm even}=99.49\mu$J/m$^2$ determined above and
a spin-spiral wavelength of 2.3nm~\cite{dmi_mnw_ferriani} 
we obtain a heat current density of $\mathscr{J}^{\rm
  Q}_{x}$=-270kW/m$^2$
for a spin spiral moving with a speed of $w$=1ms$^{-1}$.
This estimate suggests that fast domain walls moving at a speed
of the order of 100ms$^{-1}$ drive significant heat currents that
correspond
to temperature gradients of the order of 0.1K/($\mu$m). 
\section{Summary}
Magnetization dynamics drives heat currents in magnets with broken
inversion symmetry and SOI. This effect is the inverse of the thermal
spin-orbit torque. We use the Kubo linear-response formalism to
derive equations suitable to calculate the inverse thermal spin-orbit
torque (ITSOT) from first principles. We find that a ground-state
energy current associated with the Dzyaloshinskii-Moriya
interaction (DMI) is driven by magnetization dynamics and needs to be
subtracted from the linear response of the energy current in order to
extract the heat current. We show that the ground-state energy currents
obtained from the Kubo linear-response formalism can also be used 
to derive expressions for DMI and for orbital magnetization.
The ITSOT extends the picture of phenomena associated with the
coupling of spin to electrical currents and heat currents in magnets
with broken inversion symmetry and SOI.
Based on \textit{ab-initio} calculations we estimate the heat currents
driven by magnetization precession and moving spin-spirals in 
Mn/W(001) magnetic bilayers. Our estimates suggest that fast 
domain walls in magnetic bilayers drive significant heat currents.
\ack
We gratefully acknowledge computing time on the supercomputers
of J\"ulich Supercomputing Center and RWTH Aachen University
as well as financial support from the programme
SPP 1538 Spin Caloric Transport
of the Deutsche Forschungsgemeinschaft.\\

\bibliography{itsot}

\end{document}